\documentclass{elsart1p}
\usepackage{graphicx}
\usepackage{amssymb}
\begin{document}
\begin{frontmatter}
%
%
%
%
%
\title{Heavy Flavor and Jets at RHIC}
%
%

\author{Duncan Prindle for the STAR Collaboration}

\address{CENPA, University of Washington}

\begin{abstract}

We present recent results for heavy flavor and jets produced in $\rm \sqrt{s_{NN}} = 200$~GeV
p-p, d-Au, Cu-Cu and Au-Au collisions at at RHIC.
We find J/$\psi$ production is suppressed in Au-Au, but high energy J/$\psi$ are not suppressed in d-Au or Cu-Cu.
Non-photonic electrons from D and B mesons are both suppressed in Au-Au.
We study jets using two-particle 2D angular correlations as well as jet reconstruction algorithms.
We find that jets show little if any suppression, but are highly modified in central Au-Au, becoming elongated
in the $\eta$ direction and having fewer high-$\rm p_t$ particles but many more low-$\rm p_t$ particles for central Au-Au.

\end{abstract}

\begin{keyword}
RHIC\sep STAR\sep Hard scattering\sep Relativistic heavy ion collisions
%

\PACS
25.75.-q\sep 13.87.-a\sep 14.40.Pq\sep 13.20.Fe\sep 13.20.He
\end{keyword}
\end{frontmatter}

\section{Introduction}
\label{}

The RHIC is a versatile accelerator that has been used to study p-p, d-Au, Cu-Cu and Au-Au collisions.
In the present paper we are interested in the system produced in $\rm \sqrt{s_{NN}} = 200$~GeV Au-Au collisions.
Heavy flavor and jets are produced early in the collision and we measure how their properties are
modified by the medium produced in Au-Au collisions, using p-p and d-Au collisions to determine the baseline expectation.

The primary component of the STAR detector is a TPC measuring charged tracks with $\rm p_t > 0.15$~GeV/c,
covering  $2\pi$ in azimuth and $-1 < \eta < 1$ in pseudorapidity.
At larger radius is a barrel electromagnetic calorimeter, containing a shower maximum detector, which is used to measure neutral
energy and identify electrons.
The Forward Meson Spectrometer (FMS) measures $\pi^0$s for $3 < \eta_{\pi^0} < 4$.
The data acquisition system has been significantly enhanced, allowing data rates of up to 1000 events per second.
The Time of Flight (TOF) system installation was completed before the most recent run.
The TOF covers the same angular range as the TPC and increases the identification of $\pi^\pm$ and $\rm K^\pm$
to about 1.6~GeV/c and protons to 3~GeV/c as well as identifying electrons where the dE/dx crosses the
$\pi^\pm$ and $\rm K^\pm$ bands near 0.3 and 0.7~GeV/c.
An example of an analysis requiring the TOF is shown in the left panel of Fig.~\ref{electronPlots}

\section{Heavy Flavor}

Charm and bottom quarks are produced early in the collision.
Most charm and bottom quarks emerge as D and B mesons, so-called open heavy flavor,
which have $\rm c\tau$ values of hundreds of microns.
We would like to determine the c and b quark transport properties in the medium by measuring their energy loss and flow.
Some charm and bottom quarks are bound into J/$\psi$ and $\Upsilon$ and their excited states,
so-called quarkonia.
It has been predicted that quarkonia will ``melt" in a quark gluon plasma~\cite{satz}, the order of melting depending on the
binding energies and plasma temperature.

\subsection{J/$\psi$, $\Upsilon$ at $\rm \sqrt{s_{NN}} = 200$~GeV}

Quarkonia states with the lowest binding energy, the $\rm\psi(2S)$ and $\rm\Upsilon(3S)$,
melt at a lower temperature than the J/$\psi$ and $\rm\Upsilon(2S)$
which melt at a lower temperature than the $\rm\Upsilon(1S)$~\cite{thermometer}.
At RHIC the $\rm\Upsilon(1S)$ should not
melt while the J/$\psi$ and $\rm\Upsilon(2S)$ should be partially melted and the
$\rm\Upsilon(3S)$ should be almost completely melted.
This melting would result in fewer observed quarkonia than expected from extrapolation of p-p collisions.

We observe quarkonia states through the $\rm e^+e^-$ decay channel, identifying high energy electrons
by using p/E in the barrel electromagnetic calorimeter and shower max detector.
We measure suppression in A-A collisions using $\rm R_{AA} = \sigma_{AA} / N_{binary} \sigma_{pp}$,
assuming the production mechanism depends on hard scattering processes.
Normally this is measured as a function of $\rm p_t$ (so a shift in the spectrum would be
interpreted as a suppression).
For d-Au we see no strong evidence for suppression of J/$\psi$.
For Cu-Cu STAR measures the J/$\psi$ cross section above 5~GeV and sees no suppression~\cite{CuCuRAA}.
PHENIX does see a suppression at lower J/$\psi$ energies~\cite{phenixCuCuRAA}.
For Au-Au we observe a suppression with $\rm R_{AA} \approx 0.5$ for J/$\psi$ momenta below 5~GeV/c.

The $\Upsilon$ is harder to measure and in the $\rm e^+e^-$ channel it is difficult to distinguish the 1S, 2S and 3S states
so we combine them in our measurement.
We have significant signal for p-p, d-Au and Au-Au collisions at $\rm \sqrt{s_{NN}} = 200$~GeV,
finding $\rm R_{dAu} = 0.78\pm0.28\pm0.2$. 
For the most central 60\% of Au-Au collisions we measure $\rm R_{AA} = 0.78\pm0.32\pm0.22$.
The large dataset taken during the last year with less material inside the TPC and including the TOF
will reduce the uncertainties on $\rm R_{AA}$ and we may be able to tell if the suppression is consistent with the 2S and 3S
melting while the 1S stays intact.

\begin{figure}
    \includegraphics [width=1.0\textwidth]{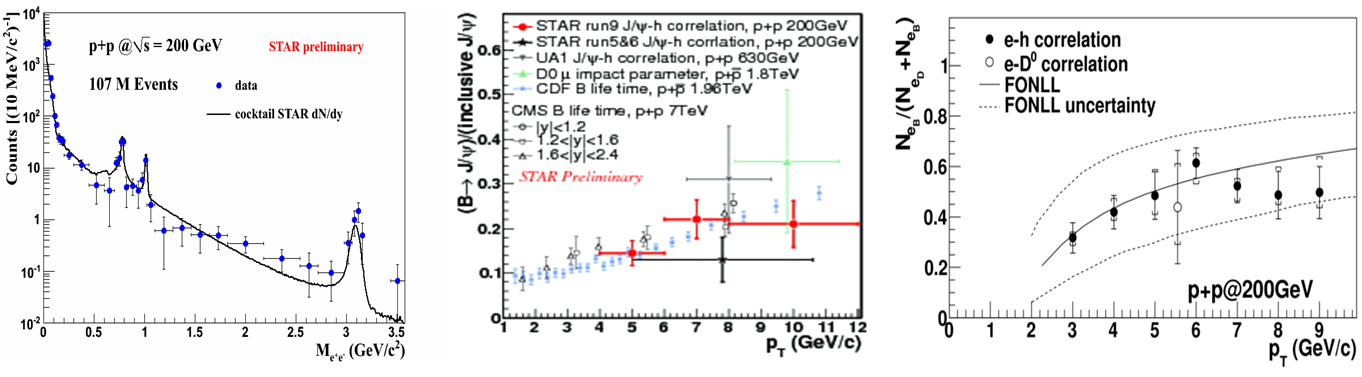}
    \caption{Left panel shows the di-electron invariant mass spectrum from 200~GeV p-p collisions.
             The peaks are the $\omega$, $\phi$ and J/$\psi$.
             The solid line is a cocktail of known contributions to the $\rm M_{e^+e^-}$ spectrum
             which describes p-p well.
             The middle panel shows the fraction of J$\psi$ that come from decay of B mesons,
             comparing p-p collisions at $\sqrt{s_{NN}} = 200$~GeV to higher energy collisions~\cite{zebo}.
             The right panel shows the fraction of non-photonic electrons from B meson decay for
             p-p collisions at $\sqrt{s_{NN}} = 200$~GeV.
    }
    \label{electronPlots}
\end{figure}

$\rm R_{AA}$ may change for reasons other than quarkonia melting.
It is possible recombination is important and/or the feed-down to J/$\psi$ from B meson decay could change.
For p-p collisions we measure this feed-down using angular correlations of hadrons with respect
to the J/$\psi$ direction~\cite{CuCuRAA}.
Using Pythia and identifying J/$\psi$s that came from B meson decay versus directly produced we get two histograms.
For B meson decays there is a same-side peak which is absent for directly produced J/$\psi$s.
We adjust the relative weights of these histograms to match the data and determine that the feed-down fraction,
shown in Fig.~\ref{electronPlots}, varies from about 10\% to over 20\% depending on J/$\psi$ momentum.
This is consistent with much higher beam energies, suggesting that once a threshold is crossed
for production of B mesons then the ratio of b to c production is constant.
If this feed-down were modified in A-A collisions $\rm R_{AA}$ of the J/$\psi$ would be affected.

\subsection{Open Heavy Flavor}

The cleanest way to measure open-heavy-flavor production would be to reconstruct D and B mesons, although this
is difficult due to small branching ratios of measurable channels.
We use the semi-leptonic decays, measuring single electrons.
Most high energy electrons observed in heavy ion collisions come from $\gamma$ conversion to $\rm e^+e^-$.
We subtract those using the small opening angle of the $\rm \gamma\rightarrow e^+e^-$, ($\rm \theta \approx m_e/E_\gamma$).
After subtraction we are left with the non-photonic electrons, a mixture of D and B meson decays.
Using Pythia we calculate electron-hadron correlations for D and B decays (the heavier B has a flatter near side correlation)
and adjust the relative amplitudes to measure $\rm r_b \equiv N_{e_B} / (N_{e_D} + N_{e_B})$ as a function of electron momentum.
Above $\rm p_t = 5$~GeV/c we find $\rm r_b\approx 0.5$ as shown in Fig.~\ref{electronPlots}~\cite{eHadron}.
It has been observed that $\rm R_{AA}$ for non-photonic electrons in $\rm \sqrt{s_{NN}} = 200$~GeV Au-Au collisions
is suppressed to about the same level as inclusive hadrons~\cite{phenixRAA}, suggesting that charm and bottom
quarks lose energy.
It had been believed that heavy quarks would have little radiative energy loss and would not be suppressed~\cite{who}.
The suppression of non-photonic $\rm R_{AA}$ in Au-Au collisions combined with $\rm r_b$ measured in p-p collisions
means that $\rm R_{AA}^{e_B} < 0.6$ at the 90\% confidence level.

\subsection{Future of Heavy Flavor in STAR}
The Time of Flight system significantly enhances STAR's particle identification capabilities and
with the large dataset recently taken we can improve all the measurements presented here.
Further into the future, we plan to install a muon telescope system outside the magnet iron, supplementing
the electron measurements with muons and resolving the $\Upsilon$ 1S, 2S and 3S states.
We are designing a heavy flavor tracker consisting of a silicon pixel detector at 2.5 and 8 cm from the
collision point, a silicon strip pad detector at 14 cm and the silicon strip detector at 22 cm, expected
to be completed for run 14.
We will be able to reconstruct $\rm D^0$  and $\rm D^\pm$ mesons, measuring $\rm R_{AA}$.
We will also measure $\rm \Lambda_C$, checking if more charm quarks end up in baryons in A-A collisions.
Identifying J/$\psi$ originating from a secondary vertex we can measure the feed-down from B mesons to J/$\psi$.
The direct measurement of D mesons will also allow us to measure $v_2$ for charm.

\subsection{$v_2$ of Heavy Particles}

We have measured $v_2$ for the light nuclei d, t and $\rm {}^3He$ produced in Au-Au collisions at $\rm \sqrt{s_{NN}} = 200$~GeV~\cite{Chitrasen Jena}.
The $v_2$ values are consistent with expectations from nucleon coalescence.
For the multiply strange particles $\phi$, $\Xi$ and $\Omega$ we also observe a $v_2$ consistent with
coalescence, this time from quarks.
By fitting a cocktail mixture to the $\pi^+\pi^-$ invariant mass distribution we have measured the $\rm p_t$ dependence
of the $\rho^0$ meson $v_2$ for the first time~\cite{Prabhat Pujahari}.
At high $\rm p_t$, above 2~GeV/c, the $\rho$ $v_2$ agrees with $\rm K^0_S$ $v_2$.
There may be a disagreement for low $\rm p_t$.
This would be interesting because if the $\rho$ is directly produced as a $\rm q\bar q$ pair it should
follow a coalescence for two quarks while if it is a recombination of two pions it should follow a coalescence for four quarks.

\section{Jets}

Jets are produced by the fragmentation of hard-scattered partons.
The scattering happens early in the collision, usually creating a pair of partons back-to-back in $\phi$
but boosted along the beam direction, dependent on the $x$s of the beam partons.
One can get some information about jets by looking at high-$\rm p_t$ particles, but because of fragmentation the
leading particle is not always a good indication of the jet energy or direction.
Better is to reconstruct jets explicitly using a jet-finding algorithm or implicitly through a technique
such as two-particle correlations.
Jet-finding algorithms group particles that came from fragmentation of a hard-scattered parton~\cite{DokJetAlgo}\cite{EllisJetAlgo}\cite{FastJet}.
They have some assumptions built in though, such as the shape of the fragmentation pattern, and if jets are sufficiently
modified in a medium these reconstruction algorithms could bias the measured jet properties.
In A-A collisions there are typically many particles within the `cone' radius that are unrelated to the
hard-scattered parton which must be subtracted.
Two-particle correlations can include all pairs of fragments and the un-related background should be subtracted naturally.
But there is some concern that other types of correlations can contribute.
In the next sections we will describe observations of jets via two-particle autocorrelations
followed by jets via explicit jet reconstruction algorithms.

\subsection{Two-Particle Correlations}
\begin{figure}
    \includegraphics [width=1.0\textwidth]{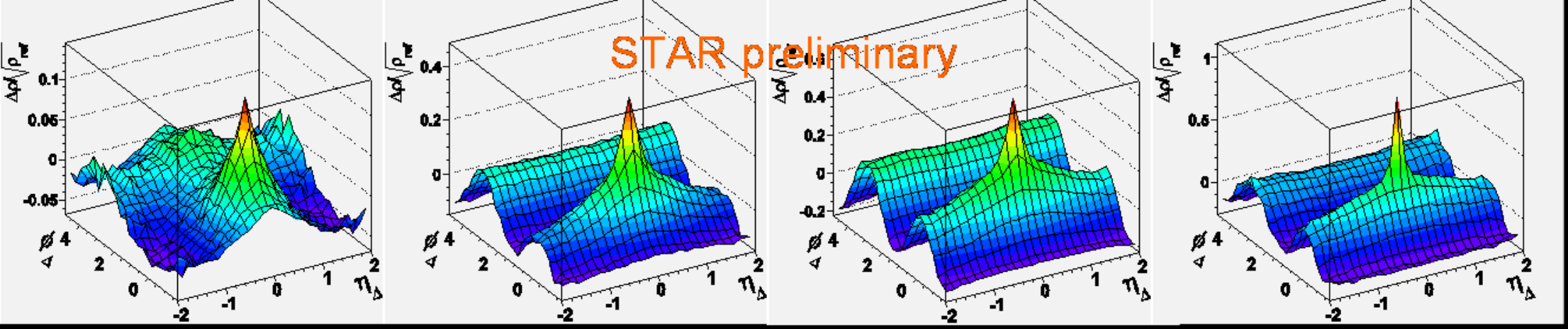}
    \caption{Samples of $(\eta_\Delta,\phi_\Delta)$ joint autocorrelations for 200~GeV Au-Au collisions.
             The left-most panel is most peripheral (84-94\%) while the right-most panel (5-9\%) is nearly central.
             The middle two panels are on either side of the sharp transition, 55-64\% and 46-55\% centralities.
    }
    \label{joint_auto-correlation}
\end{figure}

Two particle correlations measured by statistical measure
$$\rm {\Delta\rho\over\sqrt{\rho_{ref}}} \equiv {\rho(\vec{p_1},\vec{p_2})-\rho(\vec{p_1})\rho(\vec{p_2}) \over \sqrt{\rho(\vec{p_1})\rho(\vec{p_2})}}$$
are independent of $\phi_1 + \phi_2$ because of the $2\pi$ acceptance and nearly independent of
$\eta_1 + \eta_2$ within the acceptance of the STAR TPC, so we can average over those variables without loss of information.
The process of averaging over a stationary variable is called an autocorrelation.
Averaging over $\phi_1 + \phi_2$ and $\eta_1 + \eta_2$ results in a joint autocorrelation on $(\eta_\Delta,\phi_\Delta)$, where
$\eta_\Delta \equiv \eta_1-\eta_2$ and $\phi_\Delta \equiv \phi_1-\phi_2$.
To study angular correlations we can integrate over the entire observed $\rm p_{t1}$ and $\rm p_{t2}$ or a sub-range.
Examples of joint autocorrelations for selected centralities from Au-Au collisions at $\rm \sqrt{s_{NN}} = 200$~GeV
are shown in Fig.~\ref{joint_auto-correlation}.
We see evolution with centrality from a correlation structure consistent with
p-p collisions for the most peripheral collisions to a structure highly elongated on $\eta_\Delta$ for central collisions.
The peripheral correlation shape is well fit with a sharp 2D exponential (for HBT and $\rm e^+e^-$ affecting only
a few bins near (0,0)), a 2D Gaussian, an away side $\cos(\phi_\Delta-\pi)$ term, a 1D Gaussian on $\eta_\Delta$
(which is only important for peripheral centralities) and a constant offset.
For mid centralities we also need a $\cos(2\phi_\Delta)$ term.

\begin{figure}
    \includegraphics [width=1.0\textwidth]{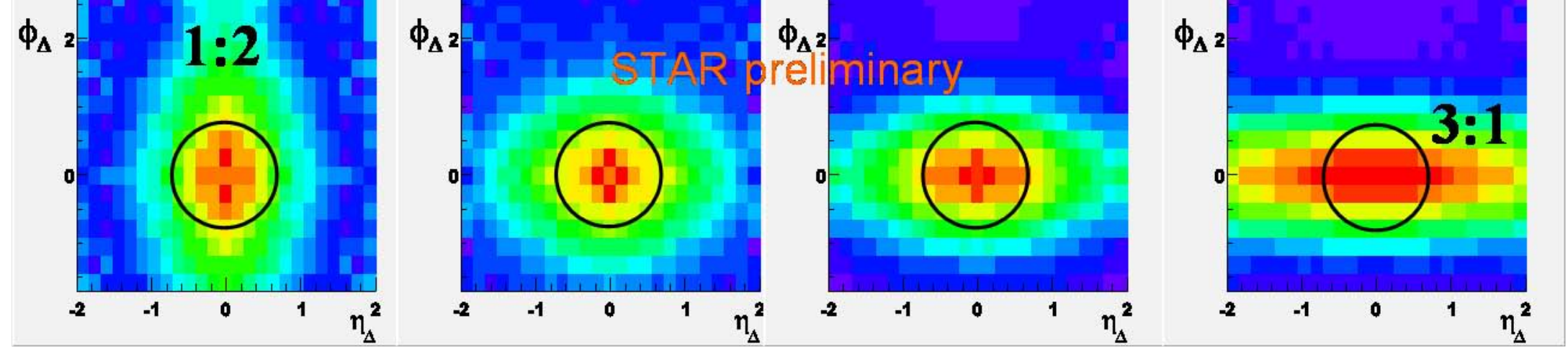}
    \caption{We have subtracted the $\rm \cos(\phi_\Delta-\pi)$, $\rm \cos(2\phi_\Delta)$ and 2D exponential
             components (determined by fitting) from the data leaving the same-side 2D Gaussian.
             The exponential affects only a few bins near (0,0).
             The away-side is flat, so we plot only the same-side, with $\phi_\Delta$ and $\eta_\Delta$ in a one-to-one
             aspect ratio.
             The circles are radius 0.7, indicating a typical jet-finding algorithm radius.
    }
    \label{joint_auto-correlation-sub}
\end{figure}

We emphasize the centrality dependence of the 2D Gaussian peak.
We fit each centrality and subtract the 2D exponential, $\rm \cos(\phi_\Delta)$ and $\rm \cos(2\phi_\Delta)$
components from the data.
The result is shown in Fig.~\ref{joint_auto-correlation-sub}.
We see that the aspect ratio of the same-side peak changes quite dramatically, from a two-to-one ratio along the $\phi_\Delta$
direction in p-p collisions to a three-to-one ratio along the $\eta_\Delta$ direction in central Au-Au collisions.
At the centrality where the aspect ratio changes from the $\phi_\Delta$ direction to the $\eta_\Delta$ direction
the 2D Gaussian amplitude increases dramatically as does the amplitude of the away-side $\cos(\phi_\Delta-\pi)$.
We interpret the same-side peak as the correlation of intra-jet fragments from hard-scattered partons
and the away side $\cos(\phi_\Delta-\pi)$ as the correlation of the inter-jet fragments.

To support our interpretation of the same-side peak arising from hard scattering we examine its $\rm p_t$
structure by selecting on $\rm p_t$ of each pair included in the correlation.
We find that for pairs with each particle around $\rm p_t = 1$~GeV/c the elongation along $\eta_\Delta$
happens at the same centrality and is just as strong as for all pairs.
Having pairs of particles around 1~GeV/c from a thermal source should be very rare.

Besides the study of jet-like structure we have also characterized the $\cos(2\phi_\Delta)$ component,
($\rm \Delta\rho[2]/\sqrt{\rho_{ref}}$), that we extract from fits to 2D angular correlations.
Previously we observed that the $\rm \Delta\rho[2]/\sqrt{\rho_{ref}}$ centrality dependence appears to be independent of
beam energy and when plotted on relative impact parameter $\rm b/b_0$ has a simple shape~\cite{energyImpact}.
The beam energy and impact parameter dependence can be written as
$$\rm\rho(b){\it v}_2^2\{2D\}(b) \equiv \Delta\rho[2]/\sqrt{\rho_{ref}} = 0.0045 R(\sqrt{s_{NN}}) \varepsilon^2 n_{bin}(b).$$
$\rm\rho(b)$ is the particle angular density, $\rm n_{bin}(b)$ is the number of N-N binary collisions (which depends on
impact parameter) and we define $v_2^2\rm\{2D\}(b)$.
The important result here is that factor $R(\sqrt{s_{NN}}) = {\log(\sqrt{s_{NN}}/13.5~GeV) \over \log(200/13.5)}$
depends only on beam energy and $\rm \varepsilon^2_{opt}(b)$ (the optical Glauber eccentricity) depends only on the initial
geometry of the collision.
We extend this to include the $\rm p_t$ dependence of $v_2^2\rm \{2D\}(b)$~\cite{ptFactor}.
We measure two-particle correlations as a function of the $\rm p_t$ of one of the particles.
Extracting the $\rm \cos(2\phi_\Delta)$ component we expect it should have a leading $\rm p_t$ dependence which we remove.
Defining
$$\rm Q(y_{t,b}) \equiv \rho_0(y_t,b) {\it v}_2^2\{2D\}(y_t,b)/p_t\over \rho_0(b) {\it v}_2^2\{2D\}(b)\left<1/p_t\right>$$
so that $\rm Q(y_{t,b})$ is unit normalized we note that all centralities for Au-Au at $\rm \sqrt{s_{NN}} = 200$~GeV
fall on a single curve, defined as $\rm Q_0(y_t)$, which happens to be a L\'evy distribution.
We can finally write the $\rm \cos(2\phi_\Delta)$ component in a form that is factorized on beam energy, centrality
and $\rm p_t$.
$$\rm \rho(y_t,b) {\it v}_2^2\{2D\}(y_t,b) = \left \{ \rho_0(b) {\it v}_2^2\{2D\}(b)\right \} \left \{ p_t\left<1/p_t \right> Q_0(y_t)\right \}.$$

\subsection{Jets at Forward Angles}
In d-Au collisions at $\rm \sqrt{s_{NN}} = 200$~GeV we can use a parton in the projectile deuteron to probe the gluon structure of the Au nucleus.
Measuring at forward angles with $3<\eta<4$ probes a range of $x$ from 0.001 to 0.003.
We measure the angular correlations of $\pi^0$ pairs.
For peripheral d-Au we find prominent peaks on both the same-side and away-side, similar to what is observed in p-p collisions.
For central d-Au collisions there is a strong suppression on the away side.
The away-side peak behavior is consistent with the back-to-back jets in peripheral collisions but single jet production
in central d-Au.
The back-to-back jets are boosted to forward angles because of the parton $x$ values.
Single jet production is expected in color glass condensate theories~\cite{cgc}, with a parton from
the deuteron scattering off the low-x gluon field of the Au nucleus.
The away-side suppression may also have explanations via other mechanisms~\cite{boris}.

\subsection{Reconstructed Jets}

\begin{figure}
    \includegraphics [width=1.0\textwidth]{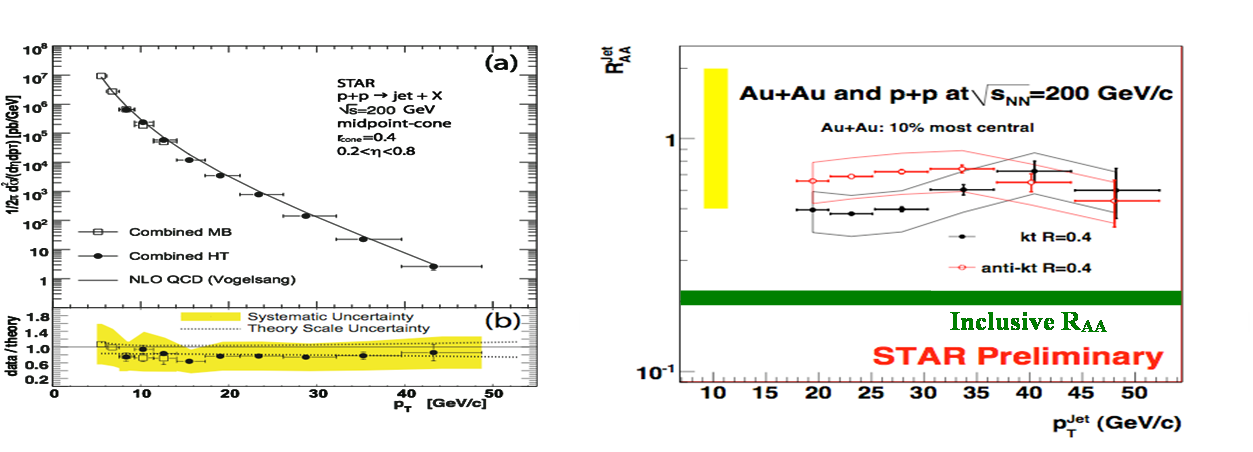}
    \caption{The left panel shows the jet cross section measured with STAR compared to a pQCD calculation for
             $\rm \sqrt{s} = 200$~GeV p-p collisions.
             There is good agreement down to jet energies of 5~GeV.
             The right panel shows $\rm R_{AA}^{jet}$ for jets reconstructed in central Au-Au collisions
             at $\rm \sqrt{s_{NN}} = 200$~GeV.
             Two different jet algorithms were used.
             The bands around the points represent the uncertainty due to background subtraction and the
             bar on the left is the uncertainty due to the energy scale of the reconstructed jet.
    }
    \label{jetCrossSection}
\end{figure}

There is a variety of jet-finding algorithms which cluster particles within some `cone'
to measure properties of the parton whose fragmentation was observed.
They don't use an actual cone, but they have a radius as a control parameter.
We calibrate our jet finding using p-p collisions~\cite{ppJet}.
The cross section, shown in Fig.~\ref{jetCrossSection}, is well described by pQCD all the
way down to jet energies of 5~GeV.
Jet fragmentation measurements are in reasonable agreement with Pythia plus Geant simulations.
Using the away-side jet to measure $\rm k_t$ in d-Au collisions we don't observe a modification of jet
properties in cold nuclear matter.

In central Au-Au collisions there will be many particles uncorrelated with the hard-scattered parton within any
cone radius large enough to measure jet properties, and these particles need to be subtracted from the reconstructed jet.
There are concerns that fluctuations in the uncorrelated particles can modify our reconstructed jet properties
but we have studied this and believe we understand the uncertainties of the background subtraction.
In Fig.~\ref{jetCrossSection} we show $\rm R_{AA}^{jet}$ and find that jets in Au-Au collisions are much less
suppressed than inclusive hadrons.
Within the uncertainties there may be no suppression.
Jets are broader in central Au-Au than in p-p.
We see this by running jet-finding algorithms with radius parameters of 0.2 and 0.4.
In p-p collisions the larger radius jet contains about twice as much energy while in central Au-Au
the larger radius jet contains four to five times as much energy.
In two-particle correlations we saw that the same-side (intra-jet) component attributed to jets is significantly broadened along
$\eta_\Delta$ in central Au-Au collisions, in general agreement with this observation.

\begin{figure}
    \includegraphics [width=1.0\textwidth]{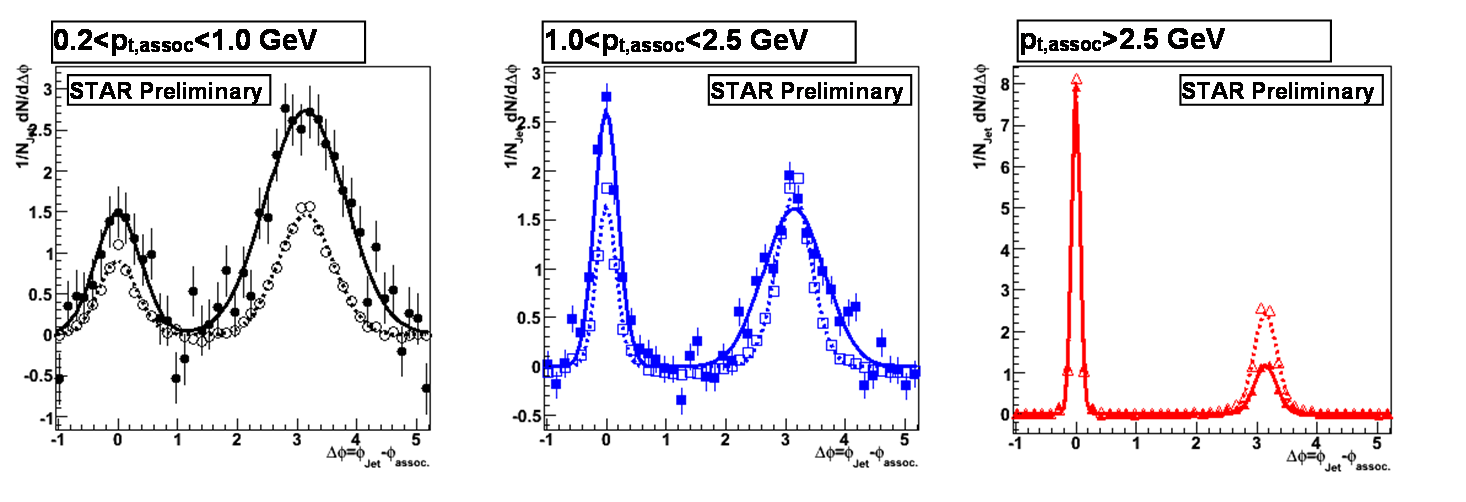}
    \caption{Jet-hadron correlations for reconstructed jets with $\rm 20 < p_{t,rec}^{jet} < 40$~GeV/c and
             three different associated hadron $\rm p_t$ ranges.
             The solid symbols are for central Au-Au collisions at $\rm \sqrt{s_{NN}} = 200$~GeV while the open
             symbols for p-p collisions at $\rm \sqrt{s} = 200$~GeV.
             This background subtraction assumed $v_2 = 0$.
    }
    \label{jetHadron}
\end{figure}

\begin{figure}
    \includegraphics [width=1.0\textwidth]{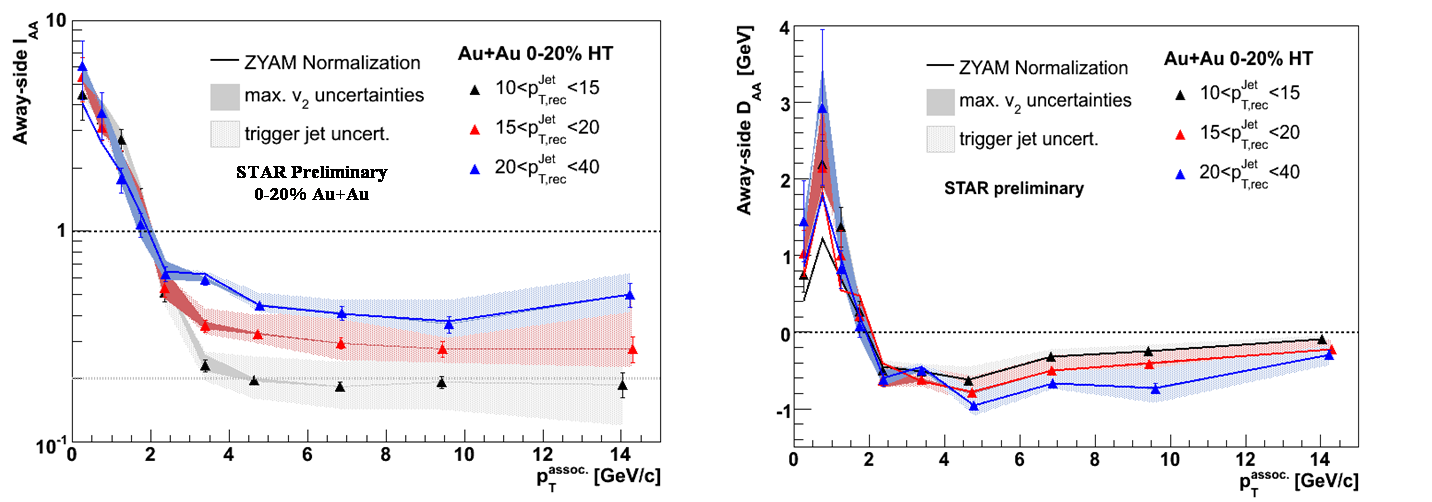}
    \caption{(Color online) The left panel is the away-side $\rm I_{AA}$ measured in central Au-Au collisions at
             $\rm \sqrt{s_{NN}} = 200$~GeV.
             For all three reconstructed jet energies there is a suppression of  associated-high-$\rm p_t$ hadrons
             on the away-side (the highest $\rm p_t$ jets having the least suppression for high $\rm p_t^{assoc}$)
             and a substantial enhancement of low-$\rm p_t$ hadrons.
             The right panel, $\rm D_{AA} \equiv p_t (Y_{AA} - Y_{pp})$, shows that the loss of associated away-side energy
             in high-$\rm p_t$ particles is largely compensated by an increase in
             energy in low-$\rm p_t$ particles.
    }
    \label{away-sideIAA_DAA}
\end{figure}

We would like to use di-jets to study medium effects.
There are possibly significant biases in our current understanding of di-jet reconstruction so for now we
look at jet-hadron correlations.
We determine the direction (and energy) of the reconstructed jet then look at $\rm \phi_{jet} - \phi_{hadron}$ for
all hadrons within a selected $\rm p_t$ range in the event.
The signal to noise ratio is much larger for jet-hadron correlations than for combinatoric hadron-hadron correlations,
so the details of the background subtraction are much less important.
In Fig.~\ref{jetHadron} we show jet-hadron correlations for three associated hadron $\rm p_t$ ranges and compare to p-p collisions.
We see that the number of low-$\rm p_t$ associated hadrons on the same side is larger for Au-Au than p-p while at high
$\rm p_t$ the numbers are equal.
On the away side there are more associated hadrons at low $\rm p_t$ in Au-Au while there are more
associated hadrons at high $\rm p_t$ in p-p.
We quantify this in Fig.~\ref{away-sideIAA_DAA}, showing that high-$\rm p_t$ hadrons are suppressed in Au-Au relative to
p-p while there is a significant enhancement of low-$\rm p_t$ hadrons.
This is similar to a two-component single-particle spectrum analysis taking the ratio of the hard components for Au-Au and p-p.
The reason the inclusive $\rm R_{AA}$ is small for low-$\rm p_t$ hadrons is because $\rm R_{AA}$ is dominated by soft particle production.
Also shown in Fig.~\ref{away-sideIAA_DAA} is the energy transport.
The away-side jet has less energy carried by high-$\rm p_t$ particles, but most if not all of this energy loss is compensated
by an increase of energy carried by lower-$\rm p_t$ particles.
We see that jets are substantially modified in the central Au-Au environment but not absorbed.

\section{Summary}

We presented measurements showing suppression of J/$\psi$ (and maybe $\Upsilon$) in Au-Au.
Non-photonic electrons from D and B meson decays are also both suppressed.
The installation of the heavy flavor tracker will greatly improve our studies of quarkonia as well
as allowing us to reconstruct D mesons to directly check for D suppression and flow.
We will be able to reconstruct $\Lambda_C$ and check if more charm goes into baryons in A-A collisions compared to p-p.

We showed that at forward rapidities the away-side jet is suppressed in d-Au collisions.
The 2D two-particle angular correlations show an evolution of the same-side jet peak from elongation in $\phi$ for
peripheral collisions to strong elongation in $\eta$ for central Au-Au collisions.
Reconstructed jets show that in central Au-Au jets are not significantly suppressed but have fewer high-$\rm p_t$ particles
and many more low-$\rm p_t$ particles.


\begin{thebibliography}{00}

  \bibitem{satz} T. Matsui and H. Satz, {\em Phys. Lett.} {\bf B178}, 416 (1986)
  \bibitem{thermometer} \`A. M\`osci and P\`eter Petreczky, {\em Phys. Rev. D} {\bf 77}, 014501 (2008)
  \bibitem{CuCuRAA} B. I. Abelev et. al., {\em Phys. Rev. C} {\bf 80} 041902(R) (2009)
  \bibitem{phenixCuCuRAA} A. Adare et al., [{\tt nucl-ex/1005.1627}]
  \bibitem{zebo} Z. Tang, (2010) [{\tt nucl-ex/1012.0233}]
  \bibitem{eHadron} M.M. Aggarwal et al., {\em Phys. Rev. Lett} {\bf 105}, 202301 (2010)
  \bibitem{phenixRAA} A. Adare et al., {\em Phys. Rev. Lett.} {\bf 98}, 172301 (2007)
  \bibitem{who} Y. L. Dokshitzer and D. E. Kharzeev, {\em Phys. Lett. B} {\bf 519}, 199 (2001), B. Z. Kopeliovich, I. K. Potashnikova and Iv\'an Schmidt, {\em Phys. Rev. C} {\bf 82}, 037901 (2010)

  \bibitem{Chitrasen Jena} C. Jena, (2011) [{\tt nucl-ex/1101.4196}]
  \bibitem{Prabhat Pujahari} P. Pujahari, (2011) [{\tt nucl-ex/1102.1286}]

  \bibitem{DokJetAlgo} S. Catani, Y. L. Dokshitzer, M. H. Seymour and B. R. Webber, \emph{Nucl. Phys. B} {\bf 406} (1993) 187
  \bibitem{EllisJetAlgo} S. D. Ellis and D. E. Soper, {\emph Phys. Rev. D} {\bf 48} (1993) 3160
  \bibitem{FastJet} Matteo Cacciari, Gavin P. Salam, {\emph Phys. Lett. B} {\bf 641} (2006) 57 [{\tt hep-ph/0512210}]
  \bibitem{energyImpact} D. Kettler, {\em Eur. Phys. J. C} {\bf 62} 175 (2009)
  \bibitem{ptFactor} D. Kettler, (2010) [{\tt nucl-ex/1011.5254}]
  \bibitem{cgc} Iancu and Venugopalan, (2003) [{\tt hep-ph/0303204}], D. Kharzeev, (2003) [{\tt hep-ph/0307037}]
  \bibitem{boris} B. Z. Kopeliovich, J. Nemchik, I. K. Potashnikova, M. B. Johnson and Ivan Schmidt, {\rm Phys. Rev. C} {\bf 72} 054606 (2005)
  \bibitem{ppJet} B. I. Abelev et al., {\em Phys. Rev. Lett.} {\bf 97}, 252001 (2006)

\end{thebibliography}
\end{document}